# Tracing technological development trajectories: A genetic knowledge persistence-based main path approach

Hyunseok Park[1,3]* and Christopher L. Magee[2,3]

[1] Department of Information System, Hanyang University, Seoul, Republic of Korea

[2] SUTD-MIT International Design Center, Massachusetts Institute of Technology (MIT), Cambridge, Massachusetts, United States

[3] Institute of Data, Systems, and Society, Massachusetts Institute of Technology (MIT), Cambridge, Massachusetts, United States

Email: Hyunseok Park (hp@hanyang.ac.kr) and Christopher L. Magee (cmagee@mit.edu)

*Corresponding author: Hyunseok Park

**Abstract**

   The aim of this paper is to propose a new method to identify main paths in a technological domain using patent citations. Previous approaches for using main path analysis have greatly improved our understanding of actual technological trajectories but nonetheless have some limitations. They have high potential to miss some dominant patents from the identified main paths; nonetheless, the high network complexity of their main paths makes qualitative tracing of trajectories problematic. The proposed method searches *backward* and *forward* paths from the high-persistence patents which are identified based on a standard genetic knowledge persistence algorithm. We tested the new method by applying it to the desalination and the solar photovoltaic domains and compared the results to output from the same domains using a prior method. The empirical results show that the proposed method overcomes the aforementioned drawbacks defining main paths that are almost 10x less complex while containing more of the relevant important knowledge than the main path networks defined by the existing method.

## 1. Introduction

   Technological progress is a major factor enabling economic growth (Schumpeter, 1934; Solow, 1957). Better understanding of technological change and innovation is essential for informing policy to enable sustainable economic and social growth. An important qualitative aspect of technological change is changes in underlying knowledge bases. Dosi's seminal work (Dosi, 1982) delineated the concepts of technological paradigm and trajectories: these have been widely used as a foundation for innovation studies (Castellacci, 2008; Christensen and Rosenbloom, 1995; Cimoli and Dosi, 1995; Frenken and Leydesdorff, 2000; Parayil, 2003; Verspagen, 2007; Von Tunzelmann et al., 2008). Tracing radical change and incremental development processes through technological trajectories provides essential insights into the evolutionary process and regularities in a technological domain (Martinelli, 2012; Verspagen, 2007). This paper contributes to the existing objective methodology for doing such studies.

In the past decade, studies on innovation have exploited patent citation networks to identify and visualize the technological trajectories from empirical data (Verspagen, 2007). A patent citation as a reference to prior art for legal purposes represents a proportion of inventive knowledge in the citing patent in originated from or already disclosed by the cited patent (Von Wartburg et al., 2005) and thus a patent citation can be considered to be a knowledge flow and sequential evolutionary path (Hall et al., 2001; Jaffe and Trajtenberg, 2002; Jaffe et al., 1993; Marco, 2007).

To reduce the complexity of a citation-based knowledge network in order to identify the most significant trajectories, Hummon and Doreian (1989) first introduced a main path algorithm and much research has applied main path analysis to technologies to investigate the patterns of technological changes (Ho et al., 2014a; Ho et al., 2014b; Huenteler et al., 2016; Martinelli, 2012; Mina et al., 2007; Verspagen, 2007). However, previous main path approaches have some limitations. First, most of them identify only one single main path. Second, they cannot show combinatorial relationships between sub-fields in a technological domain[1]. Third, traversal counts based forward searching from the starting nodes can ignore other important patents and knowledge flows (details on the limitations are described in section 2).

Therefore, the aim of this research is to propose a new main path approach to overcome the identified limitations. For this, we adopt the genetic knowledge persistence measurement (GKPM), suggested by (Martinelli and Nomaler, 2014), and differently from others, identify the Main Paths by *forward* and *backward* tracing. A GKPM quantifies how much knowledge of an invention is retained in and contributes to recent inventions based on structural and topological positions of patents in a citation network and identify high-persistence patents, whose inventive knowledge dominantly persists and contributes to recent inventions in a technological domain. Since the proposed method searches backward and forward paths from the most genetically dominant patents, it can generate multiple interconnected main paths without missing any dominantly important knowledge in the domain (we label the new method genetic backward-forward path analysis and use the acronym GBFP throughout this paper).

To test GBFP and compare it to the existing methods, we conducted empirical analyses for two technological domains, Desalination and Solar photovoltaic (PV). The results showed that GBFP identifies multiple main paths for each case with an easily recognizable number of nodes and links, including all high-persistence patents, whereas the existing approach does not identify some dominantly important patents on the identified main paths and also yields high network complexity that makes qualitatively tracing the main paths problematic.

The rest of this paper is structured as follows: Section 2 reviews the literature on the main path analysis, Section 3 describes GBFP, Section 4 presents the empirical analysis and discussion of the results, and finally conclusions are drawn in Section 5.

---

[1] We adopt the definition of a technological domain by Magee, et al., (2016): The set of artifacts that fulfill a specific generic function utilizing a particular, recognizable body of knowledge.

## 2. Main path analysis

Citation information indicates knowledge diffusing from the cited to citing documents, so knowledge flows in a citation network are used to trace evolutionary trajectories of technological or scientific knowledge. Main path analysis has been widely used to reduce the network complexity and identify the most important knowledge trajectory in a citation network. In order to identify the main sequences of citations in a large citation network, Hummon and Doreian (1989) suggested three indices, SPLC (search path link count), SPNP (search path node pair), and NPPC (node pair projection count), which calculate the 'link connectivity' based on traversal counts in search paths through the network and assign weight to each link: Batagelj (2003) later proposed one more similar index, SPC (search path count)[2]. The basic logic behind these indices is that a link (and node) included in many search paths in a citation network plays a critical role in the knowledge diffusion, so a sequence of high-weighted links (or nodes) constructs a main path.

In the initial research, main path analysis was applied to academic publications: Hummon and Doreian (1989) applied it to DNA development, Hummon and Carley (1993) applied it to the social network analysis field, Carley et al. (1993) analyzed scientific influence in the Conflict resolution field by using main path analysis. More recent research has applied the main path analysis to investigate developmental trajectories of technological fields using patent citation information: Verspagen (2007) investigated the technological trajectories of Fuel cell technology, Fontana et al. (2009) traced evolution of Local Area Network (LAN) technology, Mina et al. (2007) analyzed growth and transformation of Coronary artery disease treatment technology, Martinelli (2012) applied a main path analysis to trace the Telecommunications switching industry, Epicoco (2013) examined the long-term evolution of the Semiconductor miniaturization trajectory, Ho et al. (2014b) explored the knowledge diffusion of Membrane electrode assembly technology, and Huenteler et al. (2016) analyzed the long-term pattern of innovation and technological life-cycles in the Wind power and Solar PV fields.

Although previous approaches to main path analysis have been used for many studies, methodological limitations also have been recently discussed (Liu and Lu, 2012; Yeo et al., 2014). Here, we define the required characteristics of a main path analysis for analyzing 'technological' domains by considering the properties of a technological domain and the theoretical perspectives in innovation. First, given that a fundamental purpose of using main path analysis is to minimize the number of patents needed to realistically represent a specific technological domain, the identified main paths have to successfully contain the technologically significant patents (Fontana et al., 2009). Moreover, since technological discontinuities are usually perceived to be the significant technologies in a domain (Martinelli and Nomaler, 2014), omission of them on the main paths can make the identified main paths misleading or unreliable trajectories. For example, in the Light Emission Diode (LED) domain, a main path analysis for the domain would be judged unrealistic if the identified main paths do not contain the blue LED related patents. Second, a technological domain usually consists of several specific technological knowledge or sub-knowledge fields, so it is reasonable that developmental trajectories for a technological domain would contain multiple trajectories. For specific scientific fields or methods, such as Green chemistry (Epicoco et al., 2014), social network analysis (Carley et al., 1993) or the Hirsch Index (Liu and Lu, 2012), a singular main path might realistically trace the developmental trajectory in that the fields only focus on a narrow knowledge outcome. However, if the technological domain has many sub-technologies, for example, flexible displays, with wiring material, flexible display substrate material, thin film transistor array, semiconductor material, and so on, is shown by a singular path, this would not be useful to investigate the technological changes in the domain. Third, one major driver for technological development is the recombination of existing knowledge (Basnet and Magee, 2016; Dahlin and Behrens, 2005; Fleming, 2001; Nelson and Winter, 1982; Schilling and Green, 2011; Youn et al., 2015), therefore a main path analysis should be able to show the combinatorial relationships between the different knowledge streams which is somewhat equivalent to identifying the discontinuities as discussed above.

---
[2] Batagelj (2003) tested the performance of the search path indices (SPC, SPLC and NPPC) and concluded that these indices provide almost the same performance.

However, existing approaches are somewhat insufficient to meet the requirements. Most main path approaches produce only a singular main path (Batagelj, 2003; Hummon and Doreian, 1989; Liu and Lu, 2012; Yeo et al., 2014) and as noted above are inappropriate to analyze technological domains. Verspagen (2007) suggested a modified main path which overcomes the limitation of a singular main path and produces multiple main paths so this is the existing approach considered the baseline method in this research.

## 3. Method

In this section, genetic backward-forward path analysis (GBFP) is described in detail and its overall procedure is shown in Fig 1.

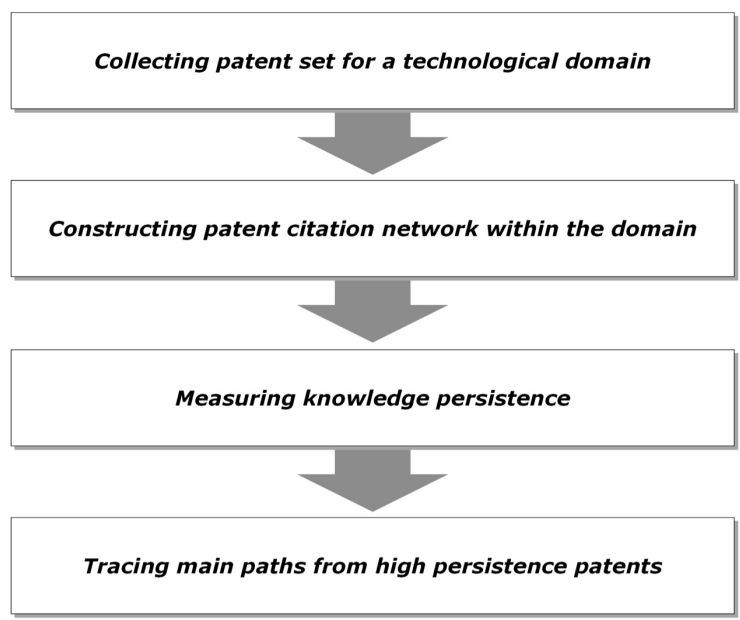

Fig. 1. The three steps in the proposed method

## 3.1. Collection of patent set

The unit of analysis of this research is a technological domain; the technological domain is defined as *'the set of artifacts that fulfill a specific generic function utilizing a particular,*

*recognizable body of knowledge'* (Magee et al., 2016). Collection of the right data is a fundamentally important step in that this can seriously affect the result. Therefore, we adopted a highly reliable patent search technique, the classification overlap method (COM) developed by (Benson and Magee, 2013, 2015), to collect highly relevant patents for a pre-defined technological domain, and downloaded the patents from www.patsnap.com.

### 3.2. Construction of patent citation network

The knowledge network of a technological domain is generated based on patent citations and, as mentioned above, the basic assumption is a patent citation represents a knowledge flow from cited patent to citing patent. This paper only considers knowledge flows within the technological domain, so patent citations occurring within the technological domain are considered (Fig 2). The cited-citing patent pairs are extracted from patent backward citation information.

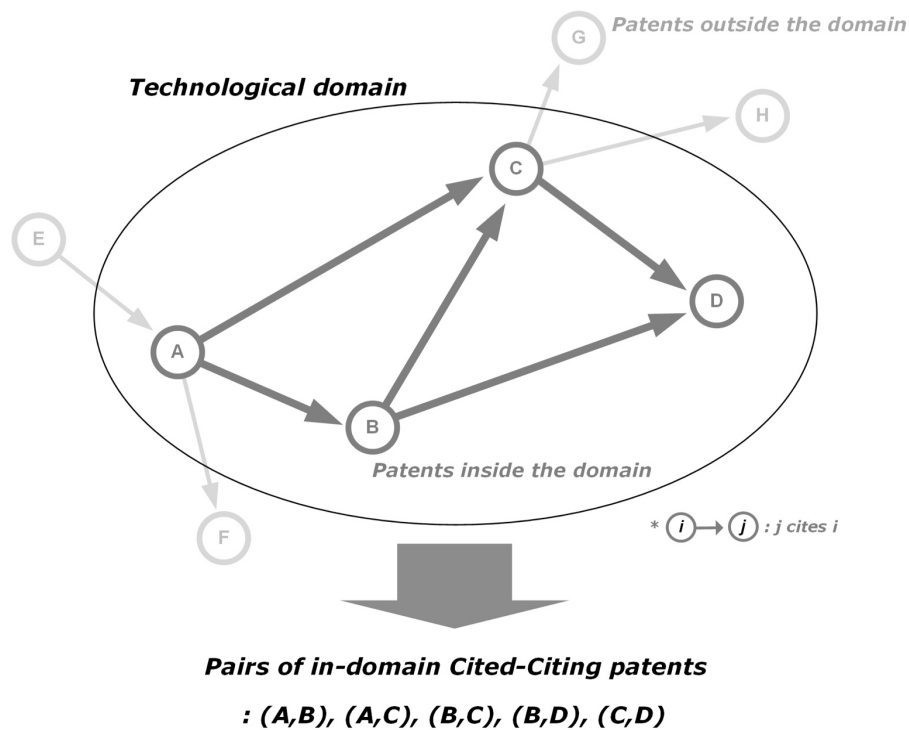

Fig 2. Cited-citing patent pairs

### 3.3. Measuring knowledge persistence

GBFP identifies main paths by a backward and forward searching from the patents dominantly important in a technological domain, i.e. high persistence patents (HPPs)[3]. Searching from the HPPs can guarantee the inclusion of all significant knowledge on the identified main paths. Both backward and forward searching can identify not only other potential main paths that might be missed by only a forward searching, but can also identify convergence structures in technological trajectories. The methodological difference of GBFP from previous approaches originates from this step. Even though the baseline approach (Verspagen, 2007) also utilizes genetic knowledge persistence, it identifies main paths by forward searching form the *startpoints* and so can miss other main paths that contain dominantly significant patents.

---
[3] The result of Martinelli and Nomaler (2014) shows that patents having high persistence value are technologically important inventions or, often, technological discontinuities in the focal technological domain.

In order to identify the HPPs, knowledge persistence of each patent is measured using the GKPM. GKPM, developed by Martinelli and Nomaler (2014), can objectively quantify the persistent knowledge of a patent by a backward mapping of the patent from all connected *endpoints*. The main concept of knowledge persistence is that a new invention is created by the recombination of the existing pieces of knowledge and so, similar to Mendelian genetic inheritance, a proportion of knowledge in a patent is incorporated in its descendant patents. Therefore, in the patent system, cited and citing patents can be interpreted as ancestors and descendants from the genetic inheritance perspective.

The procedure of GKPM is as follows (see Fig 3). First, the overall lineage structure of the technological domain is constructed by assigning each patent to a layer. The *endpoints* are identified and each patent is assigned to a layer by working backward: the *startpoints* are assigned as the first layer and then layer numbers for other patents including *endpoints* are determined. The number of layers of the domain is determined by the longest sequences of citation links from *endpoints* to *startpoints*

Second, based on the topological structure of the layer-based citation network, GKPM measures how much knowledge of a patent is inherited by recently invented patents, i.e. *endpoints*. Specifically, the proportion of the inherited knowledge of a patent to the next-generation descendant patent is calculated by 1/the number of backward citations of the next descendant patent. Therefore, knowledge persistence of a patent in the network can be calculated by the following equation:

$$KP_A = \sum_{i=1}^{n} \sum_{j=1}^{m_i} \prod_{k=1}^{l_j-1} \frac{1}{BWDCit(P_{ijk})}, \qquad (1)$$

- where $KP_A$ is knowledge persistence value of patent *A* ($P_A$),
- $n$ is the number of patents in the last layer, which are (indirectly) connected to $P_A$,
- $m_i$ is all possible backward paths from $P_i$ to $P_A$,
- $l_j$ is the number of patents on the *j*-th backward path from $P_i$ to $P_A$,
- $P_{ijk}$ is the *k*-th patent on the *j*-th backward path from $P_i$ to $P_A$, and
- $BWDCit(P_{ijk})$ is the number of backward citations of $P_{ijk}$, without considering backward citations by patents in between the first layer and layer *t-1*, when $P_A$ belongs to layer *t*.

Fig 3 shows a simple patent citation network organized from the layer perspective and gives as an example of use of equation (1) calculation of the knowledge persistence value of patent *E*. $KP_E$ is calculated by the retained knowledge in the patents J, K and L. J has three backward paths to reach at E (J→G→E, J→E and J→H→E). The number of backward citations for J is 3, for G is 2 (the backward citation to the initial patent A is ignored), and for H is 2. Therefore, $0.167 (= 0.5 \times 0.333)$ of E's knowledge is retained in J through E→G→J path, 0.33 through E→J, and $0.167 (= 0.5 \times 0.333)$ through E→H→J; the total knowledge of E retained in J is 0.667(=0.167+0.333+0.167). By the same calculation procedure, 0.5 of E's knowledge is retained in K and 0.75 in L, and so the overall knowledge persistence of E in this simple network is 1.917(=0.667+0.5+0.75).

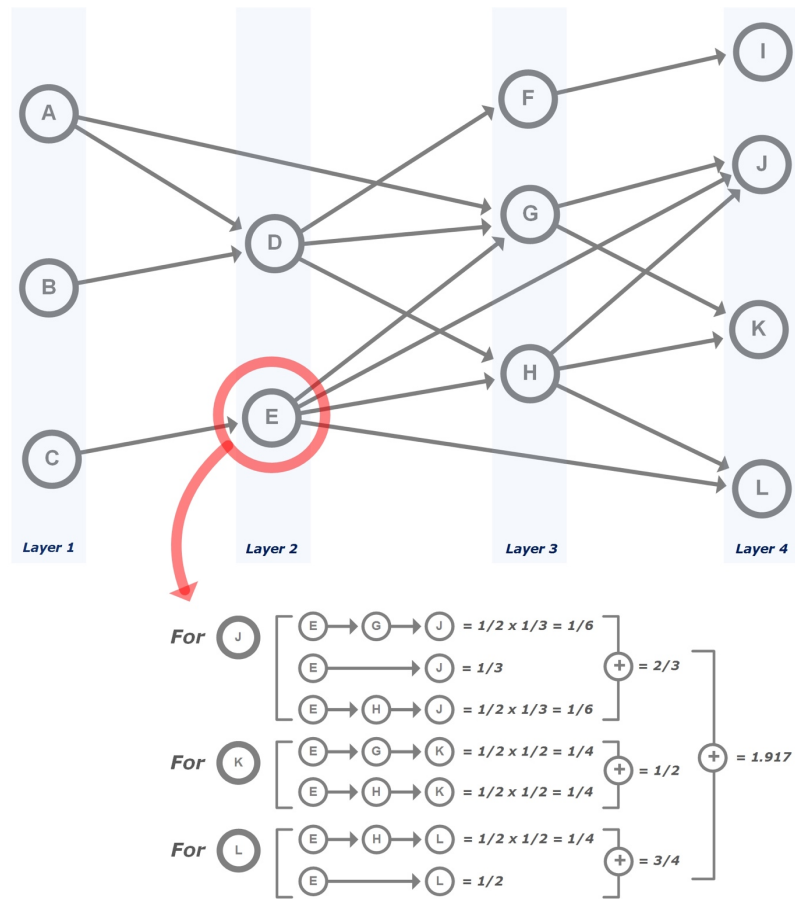

Fig 3. Measurement of knowledge persistence

### 3.4. Identification of main paths

GBFP searches paths from the patents having high knowledge persistence value. To select these HPPs, we consider two perspectives: global persistence (GP) and layer (or local) persistence (LP). GP can identify the most important patents in the domain and LP can identify the important patents in each layer: the reason that we use LP is to count the relatively recent important patents in the main paths whose overall persistence has not yet emerged. To simplify the process, we normalized the GP and LP of each patent by dividing by the maximum persistence value in the domain for GP and dividing by the maximum persistence value in each layer for LP. The cutoff value for HPPs can be set according to the desired complexity of the main paths; based on our heuristic test, GP>(0.3~0.5) or LP>(0.7~0.9) are usually most appropriate for further analysis. The lower value for the GP cutoff relative to the LP cutoff value is necessary to maximize retention of dominantly important patents in the main path network. The testing also showed a clear tradeoff between retention of important patents (favored by low cutoff values) and network complexity (favored by high cutoff values). For the rest of this paper, we use the cutoff values of GP = 0.3 and LP = 0.8 but recommend that the range suggested be investigated for completeness of analysis.

After the identification of HPPs, main paths are identified by backward and forward searching from each HPP (Fig 4): if there exist five HPPs, five backward and forward searches are performed. The basic mechanism for a backward/forward search is to choose the patent(s) having the highest GP among the directly connected cited/citing patents, therefore, any direct link between two HPPs is always chosen as a component of the main paths. The backward/forward search is finished when it reaches the *startpoint(s)/endpoint(s)* for all HPPs so all main paths are identified.

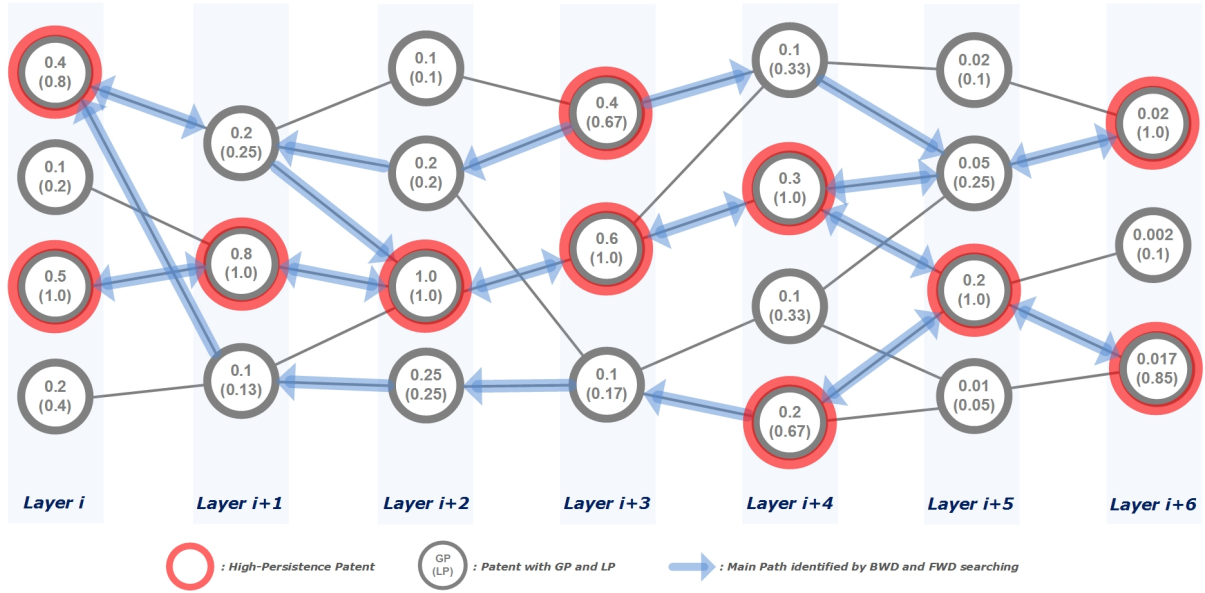

Fig 4. Searching backward and forward paths. Note: every HPP has both left and right arrows for backward and forward searches; HPP (0.2 GP and 1.0 LP) on layer *i+5* is directly connected with two HPPs on layer *i+4* and both links are chosen as main paths; if a HPP is not directly connected with other HPPs, e.g. the HPP (0.4 GP and 0.8 LP) on layer *i*, a patent which is not HPP but having the highest GP among the directly connected patents, e.g. the patent (0.2 GP and 0.25 LP) on layer *i+1*, is chosen and the further searching is continued from that patent using the same algorithm.

### 4. Empirical analysis

In this section, we present empirical analyses for two technological domains: Desalination and Solar PV. In order to show the differences of GBFP from previous methods, we compared the main paths from GBFP with the main paths from the baseline approach.

### 4.1. Solar PV

### 4.1.1. Introduction to Solar PV

Solar energy is the most abundant energy source on earth and photovoltaics have been identified as one promising type of clean energy. PV cells directly generate electricity from sunlight radiation and this PV effect was discovered over 150 years ago but has been practically developed from the 1950s (EPRI, 2009). The Solar PV domain can be broadly classified into three sub-components: solar cell, module and panel, and mounting system. The solar cell as the core component is a form of photoelectric cell which generates electricity. The PV module (or panel) is a bundle of solar cells for practical applications, and mounting systems are related to technologies to install and control a PV system. Major bottlenecks in Solar PV are conversion efficiency and costs, so the overall developmental trajectory consists of inventions that, based on the basic PV effect, adopting new materials or develop new engineering designs for alleviating these bottlenecks. Some of the patents involve sunlight concentration and hybrid structure; moreover, multi-junction cells have been developed to realize cheaper manufacturing costs while maintaining useful conversion efficiencies. Although emerging PVs, such as dye-sensitized solar cells or organic solar cells, have recently received attention, they are still at the laboratory research level (Funk and Magee, 2015) due to their currently lower performance (W/$) compared to the for now dominant PV types using crystalline silicon or amorphous silicon.

### 4.1.2. Data

The patent set for the Solar PV was obtained by COM specifically using the overlap between UPC 136 (Batteries: thermoelectric and photoelectric) and IPC H01L (Semiconductor devices; Electric solid-state devices)[4]. The number of patents in the set is 5,203, from 1976-1-1 to 2013-7-1, and the technological relevancy of the patent set is 0.85.

### 4.1.3. Result

The main paths for Solar PV drawn by the proposed method are shown in Fig 5; the graphs were drawn by using Gephi[5] (www.gephi.org) and serial numbers were given to the patents sorted by ascending order of patent numbers. The identified main paths can be broadly separated into three sub-fields: module and panel, solar cell and mounting system (Fig 5). Overall developmental trajectories are increasing conversion efficiency and reliability with lower cost by adopting new materials or new engineering designs. With the GP cutoff at 0.3 and the LP cutoff at 0.8, the main path network consists of 159 patents (58 are HPP) with 192 citations among them. Qualitatively, the main paths are relatively easy to identify in this network but larger numbers of nodes/patents make this more difficult.

---

[4] Patent search query (PatSnap): UPC:(136) AND IPC:(H01L)) AND APD:[19760101 TO 20130731]
[5] Specific Gephi's plug-in to draw Fig 5 is 'Event Graph Layout', developed by Spekkink, W., 2015. Building capacity for sustainable regional industrial systems: an event sequence analysis of developments in the Sloe Area and Canal Zone. Journal of Cleaner Production 98, 133-144.

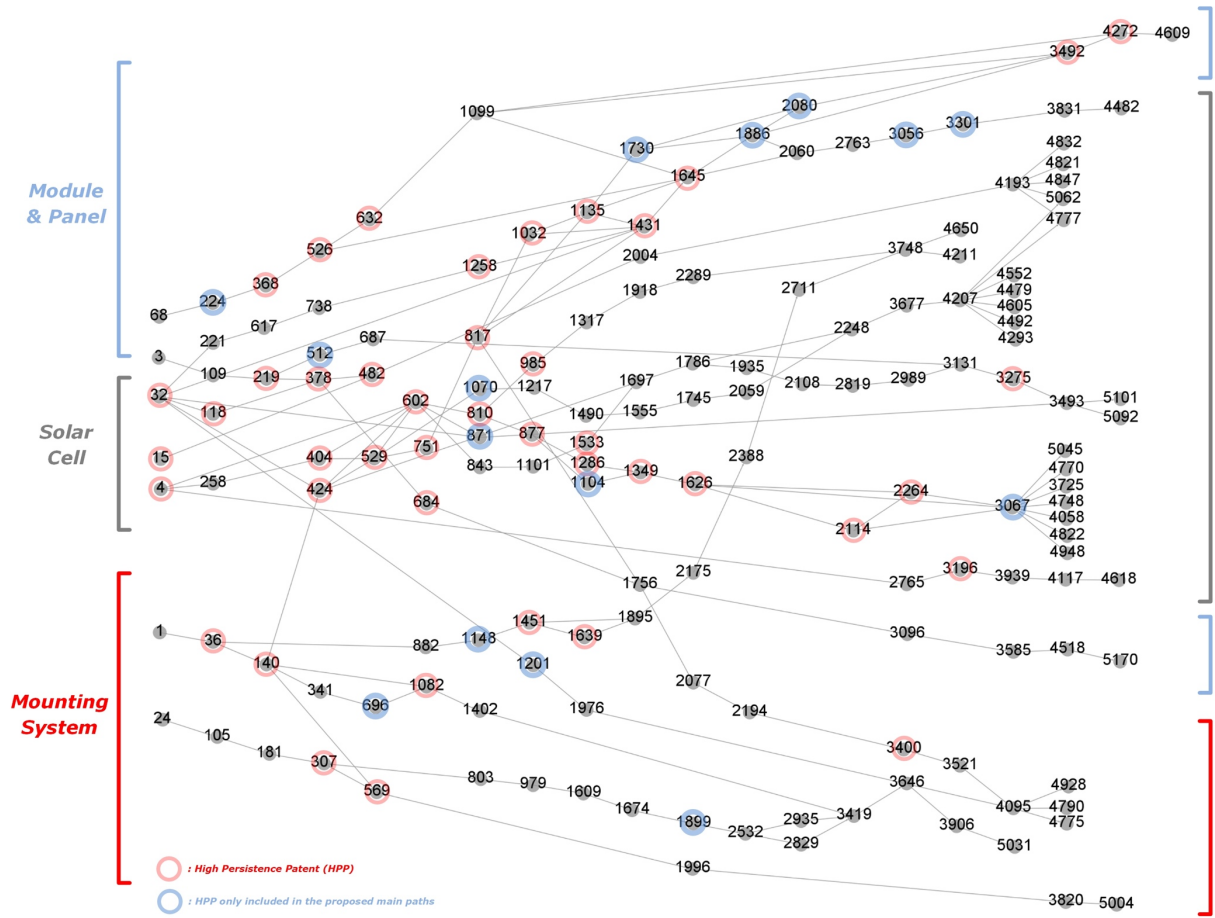

Fig 5. Main paths for Solar PV drawn by GBFP. Note: # of nodes and links are 159 and 192; # of high persistence patents (GP>0.3 or LP>0.8) are 58. (see Table 1 in Appendix for details on HPPs)

Fig 6. shows the main paths drawn by the baseline method (Verspagen, 2007). The overall network is quite large (upper left of Fig 6 has 1821 patents and 1729 citations among them which is a factor of ~9 bigger than the GBFP-based network. The larger graphs in Fig 6 were drawn (as was Fig 5) using Gephi and show some similarities to the main paths in Fig 5 with both approaches showing similar trajectories which are mainly constructed by HPPs. However, about 24% (14 patents) of the identified HPPs are only included in the main paths obtained using GBFP (in Fig 5 but not in Fig 6). Some other similarities and differences between the results from the two methods are worth noting. First, in the main paths for the Solar PV module and panel, the paths from 424 to 1645 are similar overall but the path from 1645 to 3492 and 4272 in Fig 5 and Fig 6 are somewhat different: HPPs 1730, 1886 and 2080 are included only in the path in Fig 5. Patent 1886 (US 5409549) was a new design for the solar cell module panel improving long-term reliability and cost. Specifically, the edge portions of the solar cell modules are fixed and protected by the module fasteners, the modules are not mechanically damaged, and so long-term reliability of the modules is improved. In addition, because a separated base for the work is not required, the facility and safety of the installation is improved, and cost can be reduced. Many other patents also described this invention as one representative type that is compatible with other inventions (US 6119415, US 6182404, and US7081585).

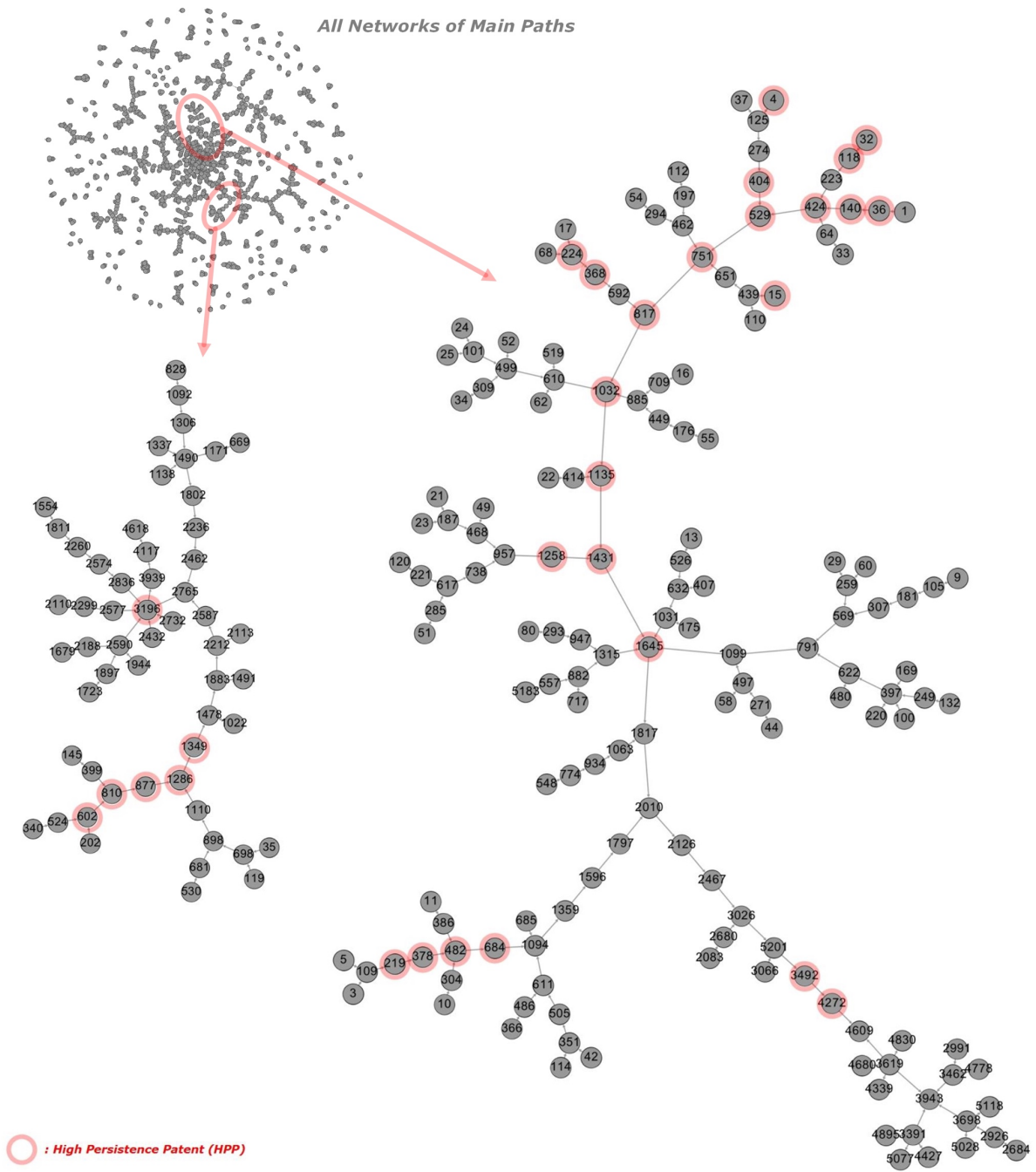

Fig 6. Main paths for Solar PV obtained by the baseline approach. Note: # of nodes and links for all main paths are 1821 and 1729; # of high persistence patents included in all main paths are 44; the left and right graphs are the sub-networks of main paths which contain more than five HPPs.

Patent 2080 (US 5589006) is a solar cell module that can be used with an air heating type passive solar system. Heat from solar energy is usually an inevitable cause of deteriorating conversion efficiency of solar cells. This solar cell module does not require an additional base, which limits the reduction in photoelectric conversion efficiency due to heat. This invention provided an important combination of solar cell modules with a passive solar system by overcoming the efficiency problem introduced by heat.

In addition, in the main paths for the solar cell, the path from 602 to 1349 is about thin film solar cell and patents in the path from both approaches are the same but the main paths in Fig 5 contain one more patent – 871. Patent 871 (US 4532537) describes a method of fabricating the photodetector comprising the light transmissive electrical contact with the textured surface on the substrate by chemical vapor deposition. This invention is an important application of surface-textured substrates for optical absorption enhancement (Yablonovitch and Cody, 1982) and many later patents introduced this as one conventional method (US 4664748, US 4880664, US 5078803 and US 5102721).

Moreover, patent 3301 (US 6784361), a relatively recent invention but not included in the main paths in Fig 6, is about amorphous silicon (a-Si) and CdS/CdTe type thin film solar cells that can provide better efficiency at elevated operation temperatures. Specifically, this includes a front electrode made of a transparent conductive oxide (TCO) and have thick intrinsic layers. Since this invention is one conventional early design of related solar cells (described in US 7846750, US 7875945, US 7888594, US 7964788, US 8203073, US 8012317, US 8022291, US 8076571, US 8133747, US 8236118, US 8334452, US 8338699, and US 8354586), leaving this patent out when tracing technological trajectories of the thin-film solar cell appears unrealistic.

### 4.2. Desalination

### 4.2.1. Introduction to desalination

The desalination domain consists of artifacts that remove salts and minerals from saline water. The potential for a global water shortage has promoted this technological domain to one of high significance for human welfare. Desalination devices can be broadly categorized as thermal or membrane-based technologies (Greenlee et al., 2009). Thermal desalination is based on water phase changes through distillation or evaporation: Multi effect distillation, multi stage flash and vapor compression distillation are the most representative technologies. Membrane desalination is based on the characteristics of semi-permeable membranes that permit water to pass through it when the pressure of feed water is greater than the osmotic pressure: reverse osmosis (RO) has been widely used for commercial purposes. Although thermal desalination accounts for a significant portion of the entire desalination market, rapid advancement of membrane desalination is apparently leading to it surpassing thermal desalinations (Greenlee et al., 2009): most patented inventions, particularly in our patent set, are related to the membrane desalination.

### 4.2.2. Data

The patent set for the Desalination technology is obtained by COM. The specific overlap used was between UPC 210 (Liquid purification or separation) and IPC C02F (Treatment of water, waste water, sewage, or sludge) or B01D (Separation). Since this classification overlap contains patents related to water treatment or purification, we added a keyword search query[6] to isolate only patents relevant to the desalination technology. The number of patents in the set is 3,634, from 1976-1-1 to 2013-7-1, and the technological relevancy of the patent set is 0.87.

### 4.2.3. Result

The identified main paths for Desalination are shown in Fig 7 for GBFP and in Fig 8 for the baseline method (Verspagen, 2007). The networks were again drawn by using Gephi and serial numbers are given to the patents sorted by ascending order of patent numbers. As with Solar PV, the baseline main path network is much larger (1744 patents with 1508 citation links among them) than the network from GBFP (115 patents with 134 citation links among them). Although sections of the baseline network can be isolated as shown in Fig 8, the added complexity makes this main path network less amenable to qualitative analysis than is the GBFP-based network shown in Fig 7.

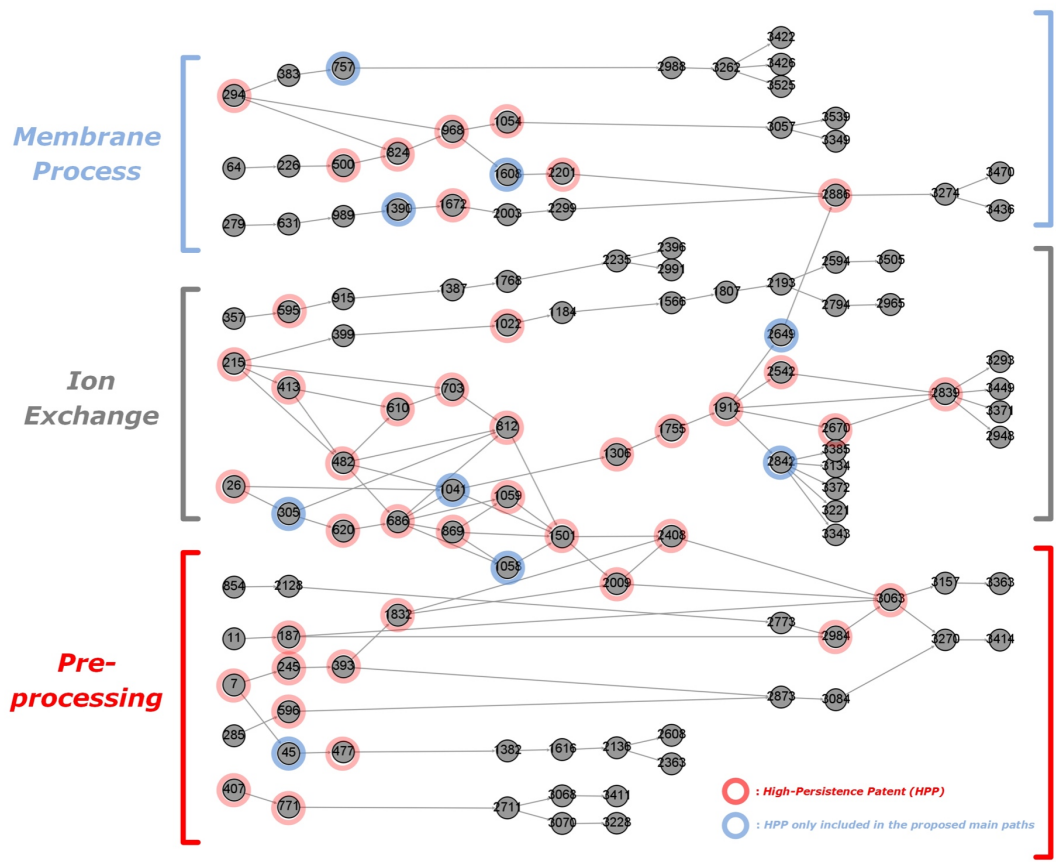

Fig 7. Main paths for Desalination technology drawn by GBFP. Note: # of nodes and links are 115 and 134; # of high persistence patents (GP>0.3 or LP>0.8) are 50. (see Table 2 in Appendix for details on HPPs)

---

[6] Patent search query (PatSnap): ((TTL:(water and (desalinat or sea or salin or salt or brackish)) OR ABST:(water and (desalinat or sea or salin or salt or brackish)) OR CLMS:(water and (desalinat or sea or salin or salt or brackish))) AND UPC:(210) AND IPC:(C02F or B01D)) AND APD:[19760101 TO 20130731]

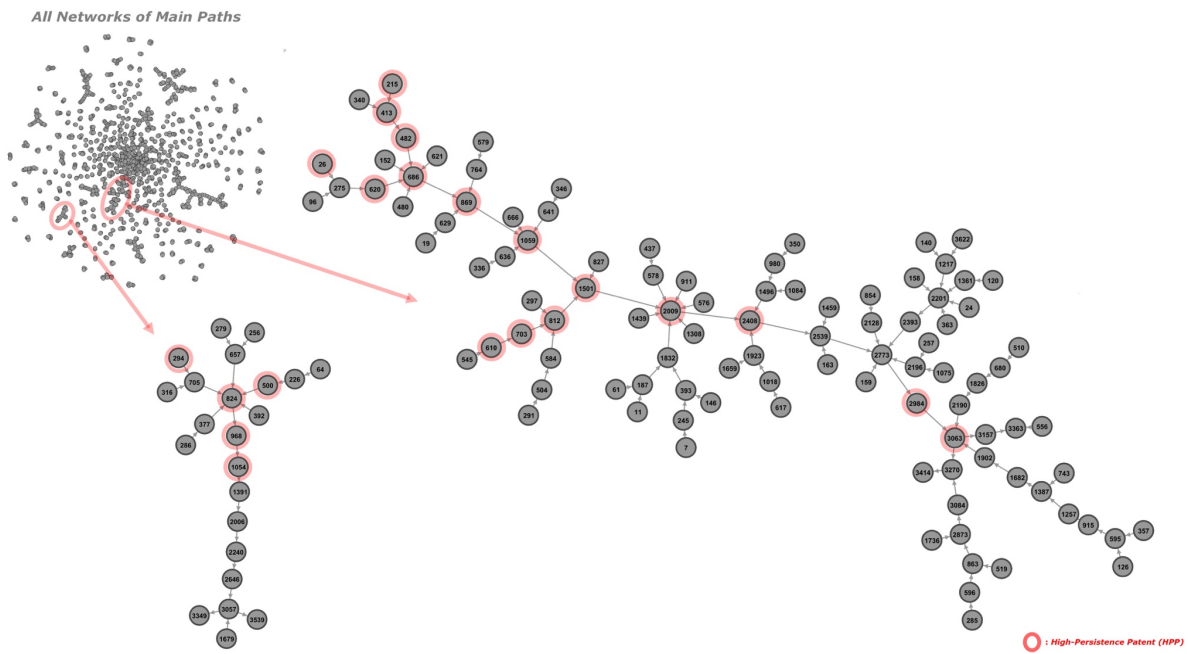

Fig 8. Main paths for Desalination technology obtained by the baseline approach. Note: # of nodes and links are 1774 and 1508; # of high persistence patents included in the all main paths are 41; the left and right graphs are the sub-networks of main paths which contain more than five HPPs.

The identified main paths can be overall classified into three sub-fields: reverse, or forward, osmosis based membrane process, ion-exchange including water softening and electrodialysis, and preprocess techniques, usually, for membrane processes including dewatering and precipitation. The major outcomes of the developmental trajectories are increasing desalination efficiency at lower energy consumption, i.e. cost apparently arising from hybrid use of different methods or new materials, e.g. nanofiltration, and new energy sources, e.g. solar energy. Both the proposed and baseline methods' main paths show overall similar trajectories.

However, similar to the PV case, the network obtained from the baseline approach – despite its much larger size – does not contain about 18% of HPPs (9 out of 50) present in the GBFP-based network. As before, qualitative analysis indicates that some of them seem to be significant knowledge inputs in tracing developmental trajectories. First, the patent 757 (US 4704324) is an invention describing a method to form a semi-permeable membrane. Specifically, a permselective discriminating layer of the membrane prepared by reaction of onium compounds with nucleophilic compounds can greatly enhance the flux of membranes and thus increase desalination efficiency of RO systems. Since many later patents adopted its inventive knowledge (US 4812238, US 4888116, US 5238747 and US 5310581), patent 757's position on the main paths appears important for further evolution of the trajectory on the flux of membranes (Fig 7). Missing this patent suggests a main path that is probably not a reliable representation of reality.

Second, the patent 1041 (US 4936987) disclosed a water soluble polymer that can prevent the precipitation or crystallization of scale-forming salts of alkaline earth metal cations. Although later inventions pointed out that large dosages of the polymers are required (US 5256302 and US 5393456), this invention is a relatively early technology for scale and corrosion inhibitors which have made substantial contribution to later relevant

developments (US 5182028, US 5259974, US 5322636, US 5338477, US 5358642, US 5284590, US 6333005, US 6355214 and US 6646082).

Third, the patent 1608 (US 5458781) is a method to separate the monovalent anion bromide from sea and brackish waters by using a combination of RO and nanofiltration (NF) membranes. Even though this patent has not been widely cited by the later patents and its major purpose is not exactly for producing potable water from sea or brackish waters, the inventive knowledge of this invention had a significant impact on further developments in combinations of RO and NF membranes; many of the patents that cite the patent 1608 are directly affected by this invention (US 6190556, US 7144511, US 8366924, and US 9205383).

## 5. Conclusion

Main path analysis has been widely adopted as a useful tool to empirically trace technological trajectories. Ideally, a main path analysis would be able to reduce network complexity for effective qualitative investigation of the technological trajectories while not eliminating dominantly important knowledge. Our results for the solar PV and the desalination domains clearly show that our proposed new approach (genetic backward-forward main path analysis (or GBFP) is a significant step towards this ideal compared to the existing approaches represented by the baseline approach. In the two cases, the GBFP networks are about 10 times *smaller* and also contain about 20% *more* of the dominant genetic knowledge in the domain. Defining less complex networks that nonetheless contain more of the important knowledge makes simultaneous progress along both key dimensions toward a better method. The GBFP does this while adopting the best practice of the baseline approach – the genetic knowledge definition as first defined by Verspagen (2007). This is done through adopting the persistence measurement (GKPM) to first identify the patents having the dominantly significant knowledge bases, i.e. HPPs, in the technological domain. Then, main paths are identified by backward and forward searching from the identified HPPs.

To verify the usefulness of the proposed method, we conducted empirical analyses for the solar PV and desalination technological domains and compared our method with the existing approach. The empirical results show that major technological trajectories on both main paths are quite similar and they are overall represented by HPPs. Most HPPs on the main paths are actually critical for dominantly important knowledge streams for the technological domains. In regard to combinatorial relationships, both approaches appropriately identify the important combinations of knowledge streams (e.g. two converging main paths onto patent 424 in Solar PV). Even though main paths from the baseline method show much more combinatorial relationships, it is apparently due to the baseline method identifying much larger networks, and many of the converging paths seem to be noise, also due to the large network size. Our qualitative analysis also found that some HPPs only included in the main paths obtained by GBFP involve significant domain knowledge and should be contained on realistic main paths. In addition, even though many of previous studies using the existing approach focused on the largest sub-network of main paths to trace technological trajectories, the empirical result shows that the largest, or even second or third largest, sub-network of main paths is not always the most important main path. This means that most of large sub-networks need to be analyzed to uncover the appropriate trajectories. Therefore, given that a major reason to use main path analysis is to reduce the network complexity, the high network complexity of the baseline main paths is not a negligible issue in that the size of the main paths, shown in Fig 6 and Fig 8, is still too large for qualitative analysis.

However, there exist some issues that would be improved in further research. First, we adopt GKPM to identify HPPs in a patent citation network. GKPM considers that a citing patent receives same proportion of knowledge from all cited patents, but the proportions in knowledge inheritance for each citation might be different. Our empirical analysis and Martinelli and Nomaler (2014)'s research show that this weighting issue is negligible when the size of a citation network is relatively large, but it might cause a reliability problem when the network size is too small. Therefore, the improvement of this weighting algorithm may increase quality and reliability of the main paths. Second, we use a concept of layer persistence (LP) to identify the patents which are recently invented but have the potential to be dominantly important. Nonetheless, the identified main paths still have a relatively low number of recent HPPs. Therefore, development of other criteria using a text mining technique to identify the recently invented but technologically important patents may provide additional value.

## Acknowledgements

The authors acknowledge the funding support for this work received from the SUTD-MIT International Design Center and the Hanyang University (HY-2016).

## Appendix

Table 1. High persistence patents in Solar PV

| Patent Number | Serial Number | Layer | Application Year | Persistence | GP | LP | Number of Indomain Forward Citations | Title |
|---|---|---|---|---|---|---|---|---|
| US4017332 | 4 | 1 | 1977 | 41.1465 | 0.365 | 0.458 | 31 | Solar cells employing stacked opposite conductivity layers |
| US4042418 | 15 | 1 | 1977 | 33.8803 | 0.300 | 0.377 | 19 | Photovoltaic device and method of making same |
| US4064521 | 32 | 1 | 1977 | 89.9046 | 0.797 | 1.000 | 75 | Semiconductor device having a body of amorphous silicon |
| US4070206 | 36 | 2 | 1978 | 59.58 | 0.528 | 1.000 | 16 | Polycrystalline or amorphous semiconductor photovoltaic device having improved collection efficiency |
| US4126150 | 118 | 2 | 1978 | 41.6985 | 0.369 | 0.700 | 10 | Photovoltaic device having increased absorption efficiency |
| US4133698 | 140 | 3 | 1979 | 55.545 | 0.492 | 0.893 | 16 | Tandem junction solar cell |
| US4166880 | 219 | 3 | 1979 | 62.1917 | 0.551 | 1.000 | 3 | Solar energy device |
| US4167644 | 224 | 2 | 1979 | 38.3252 | 0.340 | 0.643 | 7 | Solar cell module |
| US4209347 | 307 | 4 | 1980 | 48.7293 | 0.432 | 0.656 | 9 | Mounting for solar cell |
| US4239555 | 368 | 3 | 1980 | 57.3501 | 0.508 | 0.922 | 25 | Encapsulated solar cell array |
| US4245386 | 378 | 4 | 1981 | 74.2996 | 0.658 | 1.000 | 13 | Method of manufacturing a solar cell battery |
| US4255211 | 404 | 4 | 1981 | 42.3455 | 0.375 | 0.570 | 28 | Multilayer photovoltaic solar cell with semiconductor layer at shorting junction interface |
| US4272641 | 424 | 4 | 1981 | 58.2041 | 0.516 | 0.783 | 29 | Tandem junction amorphous silicon solar cells |
| US4315096 | 482 | 5 | 1982 | 82.886 | 0.734 | 0.748 | 25 | Integrated array of photovoltaic cells having minimized shorting losses |
| US4328390 | 512 | 4 | 1982 | 52.6459 | 0.466 | 0.709 | 11 | Thin film photovoltaic cell |
| US4336413 | 526 | 4 | 1982 | 37.0537 | 0.328 | 0.499 | 14 | Solar panels |
| US4338480 | 529 | 5 | 1982 | 110.8101 | 0.982 | 1.000 | 14 | Stacked multijunction photovoltaic converters |
| US4361717 | 569 | 5 | 1982 | 51.0159 | 0.452 | 0.460 | 18 | Fluid cooled solar powered photovoltaic cell |
| US4377723 | 602 | 6 | 1983 | 112.874 | 1.000 | 1.000 | 23 | High efficiency thin-film multiple-gap photovoltaic device |
| US4392009 | 632 | 5 | 1983 | 34.8606 | 0.309 | 0.315 | 10 | Solar power module |
| US4419530 | 684 | 6 | 1983 | 36.6613 | 0.325 | 0.325 | 18 | Solar cell and method for producing same |
| US4427839 | 696 | 5 | 1984 | 46.6267 | 0.413 | 0.421 | 20 | Faceted low absorptance solar cell |
| US4461922 | 751 | 6 | 1984 | 86.9645 | 0.770 | 0.770 | 36 | Solar cell module |
| US4496788 | 810 | 7 | 1985 | 85.2419 | 0.755 | 1.000 | 14 | Photovoltaic device |
| US4499658 | 817 | 7 | 1985 | 57.5016 | 0.509 | 0.675 | 36 | Solar cell laminates |
| US4532537 | 871 | 7 | 1985 | 43.0718 | 0.382 | 0.505 | 15 | Photodetector with enhanced light absorption |
| US4536607 | 877 | 8 | 1985 | 98.4017 | 0.872 | 1.000 | 22 | Photovoltaic tandem cell |
| US4609771 | 985 | 8 | 1986 | 39.1756 | 0.347 | 0.398 | 7 | Tandem junction solar cell devices incorporating improved microcrystalline p-doped semiconductor alloy material |
| US4636578 | 1032 | 8 | 1987 | 73.8888 | 0.655 | 0.751 | 18 | Photocell assembly |

| Patent Number | Serial Number | Layer | Application Year | Persistence | GP | LP | Number of Indomain Forward Citations | Title |
|---|---|---|---|---|---|---|---|---|
| US4658086 | 1070 | 7 | 1987 | 34.5753 | 0.306 | 0.406 | 11 | Photovoltaic cell package assembly for mechanically stacked photovoltaic cells |
| US4665277 | 1082 | 6 | 1987 | 40.1487 | 0.356 | 0.356 | 12 | Floating emitter solar cell |
| US4680422 | 1104 | 9 | 1987 | 43.937 | 0.389 | 0.419 | 16 | Two-terminal, thin film, tandem solar cells |
| US4692557 | 1135 | 9 | 1987 | 104.9566 | 0.930 | 1.000 | 30 | Encapsulated solar cell assemblage and method of making |
| US4698455 | 1148 | 7 | 1987 | 62.9695 | 0.558 | 0.739 | 3 | Solar cell with improved electrical contacts |
| US4732621 | 1201 | 8 | 1988 | 37.1384 | 0.329 | 0.377 | 8 | Method for producing a transparent conductive oxide layer and a photovoltaic device including such a layer |
| US4773944 | 1258 | 7 | 1988 | 38.8053 | 0.344 | 0.455 | 23 | Large area, low voltage, high current photovoltaic modules and method of fabricating same |
| US4795501 | 1286 | 9 | 1989 | 44.4999 | 0.394 | 0.424 | 10 | Single crystal, heteroepitaxial, GaAlAs/CuInSe.sub.2 tandem solar cell and method of manufacture |
| US4867801 | 1349 | 10 | 1989 | 48.7505 | 0.432 | 0.962 | 7 | Triple-junction heteroepitaxial AlGa/CuInSe.sub.2 tandem solar cell and method of manufacture |
| US4953577 | 1431 | 10 | 1990 | 50.6626 | 0.449 | 1.000 | 13 | Spray encapsulation of photovoltaic modules |
| US4971632 | 1451 | 8 | 1990 | 60.3181 | 0.534 | 0.613 | 9 | Miniature thermoelectric converters |
| US5057163 | 1533 | 9 | 1991 | 46.7949 | 0.415 | 0.446 | 18 | Deposited-silicon film solar cell |
| US5141564 | 1626 | 11 | 1992 | 44.7518 | 0.396 | 0.684 | 23 | Mixed ternary heterojunction solar cell |
| US5156688 | 1639 | 9 | 1992 | 34.0272 | 0.301 | 0.324 | 9 | Thermoelectric device |
| US5164020 | 1645 | 11 | 1992 | 65.3904 | 0.579 | 1.000 | 27 | Solar panel |
| US5252141 | 1730 | 10 | 1993 | 46.6132 | 0.413 | 0.920 | 17 | Modular solar cell with protective member |
| US5409549 | 1886 | 12 | 1995 | 46.6178 | 0.413 | 1.000 | 35 | Solar cell module panel |
| US5421909 | 1899 | 11 | 1995 | 35.1994 | 0.312 | 0.538 | 16 | Photovoltaic conversion device |
| US5589006 | 2080 | 13 | 1996 | 46.5557 | 0.412 | 1.000 | 34 | Solar battery module and passive solar system using same |
| US5626688 | 2114 | 14 | 1997 | 28.136 | 0.249 | 1.000 | 39 | Solar cell with chalcopyrite absorber layer |
| US5858121 | 2264 | 15 | 1999 | 15.9302 | 0.141 | 1.000 | 7 | Thin film solar cell and method for manufacturing the same |
| US6525264 | 3056 | 15 | 2003 | 14.3095 | 0.127 | 0.898 | 6 | Thin-film solar cell module |
| US6534704 | 3067 | 17 | 2003 | 7 | 0.062 | 1.000 | 7 | Solar cell |
| US6660928 | 3196 | 16 | 2003 | 12.5 | 0.111 | 1.000 | 12 | Multi-junction photovoltaic cell |
| US6750394 | 3275 | 17 | 2004 | 6.5 | 0.058 | 0.929 | 8 | Thin-film solar cell and its manufacturing method |
| US6784361 | 3301 | 16 | 2004 | 12.5 | 0.111 | 1.000 | 14 | Amorphous silicon photovoltaic devices |
| US6940008 | 3400 | 15 | 2005 | 14.0341 | 0.124 | 0.881 | 12 | Semiconductor device, solar cell module, and methods for their dismantlement |
| US7178295 | 3492 | 18 | 2007 | 4 | 0.035 | 1.000 | 4 | Shingle assembly |
| US8065844 | 4272 | 19 | 2011 | 1 | 0.009 | 1.000 | 1 | Ballasted photovoltaic module and module arrays |

## Table 2. High persistence patents in Desalination

| Patent Number | Serial Number | Layer | Application Year | Persistence | GP | LP | Number of Indomain Forward Citations | Title |
|---|---|---|---|---|---|---|---|---|
| US4277344 | 294 | 1 | 1979 | 20.6939 | 0.57373 | 1 | 57 | Interfacially synthesized reverse osmosis membrane |
| US4640793 | 686 | 4 | 1985 | 36.0691 | 1 | 1 | 43 | Synergistic scale and corrosion inhibiting admixtures containing carboxylic acid/sulfonic acid polymers |
| US4543190 | 595 | 2 | 1984 | 13.072 | 0.36242 | 0.53097 | 20 | Processing methods for the oxidation of organics in supercritical water |
| US5925255 | 2009 | 8 | 1997 | 26.8823 | 0.7453 | 1 | 14 | Method and apparatus for high efficiency reverse osmosis operation |
| US6537456 | 2408 | 9 | 1999 | 17.875 | 0.49558 | 1 | 15 | Method and apparatus for high efficiency reverse osmosis operation |
| US4209398 | 215 | 1 | 1978 | 19.0923 | 0.52932 | 0.9226 | 34 | Water treating process |
| US6190556 | 2201 | 7 | 1998 | 14.5361 | 0.40301 | 0.44296 | 13 | Desalination method and apparatus utilizing nanofiltration and reverse osmosis membranes |
| US4545862 | 596 | 2 | 1982 | 13.5392 | 0.37537 | 0.54995 | 9 | Desalination device and process |
| US4188291 | 187 | 2 | 1978 | 12.8157 | 0.35531 | 0.52056 | 13 | Treatment of industrial waste water |
| US4872984 | 968 | 5 | 1988 | 28.1749 | 0.78114 | 0.95811 | 31 | Interfacially synthesized reverse osmosis membrane containing an amine salt and processes for preparing the same |
| US4720346 | 771 | 2 | 1986 | 12.7835 | 0.35442 | 0.51925 | 25 | Flocculation processes |
| US4936987 | 1041 | 5 | 1988 | 17.989 | 0.49874 | 0.61173 | 13 | Synergistic scale and corrosion inhibiting admixtures containing carboxylic acid/sulfonic acid polymers |
| US4366063 | 393 | 3 | 1981 | 19.2097 | 0.53258 | 0.59252 | 16 | Process and apparatus for recovering usable water and other materials from oil field mud/waste pits |
| US4704324 | 757 | 3 | 1985 | 11.4309 | 0.31692 | 0.35258 | 12 | Semi-permeable membranes prepared via reaction of cationic groups with nucleophilic groups |
| US4761234 | 824 | 4 | 1986 | 27.4843 | 0.76199 | 0.76199 | 17 | Interfacially synthesized reverse osmosis membrane |
| US5520816 | 1672 | 5 | 1994 | 13.3378 | 0.36978 | 0.45356 | 6 | Zero waste effluent desalination system |
| US4443340 | 482 | 3 | 1983 | 32.4204 | 0.89884 | 1 | 7 | Control of iron induced fouling in water systems |
| US5695643 | 1832 | 4 | 1993 | 16.5296 | 0.45828 | 0.45828 | 9 | Process for brine disposal |
| US5254257 | 1390 | 4 | 1993 | 13.4724 | 0.37352 | 0.37352 | 9 | Reclaiming of spent brine |
| US4752443 | 812 | 6 | 1986 | 22.4102 | 0.62131 | 1 | 6 | Cooling water corrosion inhibition method |
| US5358640 | 1501 | 7 | 1993 | 32.816 | 0.90981 | 1 | 7 | Method for inhibiting scale formation and/or dispersing iron in reverse osmosis systems |
| US4948507 | 1054 | 6 | 1989 | 16.6078 | 0.46044 | 0.74108 | 21 | Interfacially synthesized reverse osmosis membrane containing an amine salt and processes for preparing the same |
| US5792369 | 1912 | 10 | 1996 | 14.6667 | 0.40663 | 1 | 10 | Apparatus and processes for non-chemical plasma ion disinfection of water |
| US4067806 | 45 | 2 | 1976 | 16.6775 | 0.46237 | 0.67742 | 15 | Formulation and application of compositions for the detackification of paint spray booth wastes |
| US4288327 | 305 | 2 | 1979 | 11.2254 | 0.31122 | 0.45596 | 18 | Copolymers for the control of the formation and deposition of materials in aqueous mediums |
| US7744761 | 3063 | 13 | 2008 | 4.6667 | 0.12938 | 1 | 6 | Desalination methods and systems that include carbonate compound precipitation |
| US4560481 | 610 | 4 | 1984 | 25.8945 | 0.71791 | 0.71791 | 22 | Method of controlling iron induced fouling in water systems |
| US7595001 | 2984 | 12 | 2003 | 4.5 | 0.12476 | 1 | 5 | Process for the treatment of saline water |
| US4048066 | 26 | 1 | 1976 | 12.1301 | 0.3363 | 0.58617 | 14 | Method of inhibiting scale |
| US4913823 | 1022 | 6 | 1988 | 11.6237 | 0.32226 | 0.51868 | 10 | Process for dissolving and removing scale from aqueous systems |
| US4440647 | 477 | 3 | 1983 | 17.2562 | 0.47842 | 0.53226 | 14 | Paint spray booth detackification composition and method |
| US4659482 | 703 | 5 | 1986 | 24.0297 | 0.66621 | 0.81715 | 21 | Water treatment polymers and methods of use thereof |
| US4382864 | 407 | 1 | 1981 | 12.3547 | 0.34253 | 0.59702 | 7 | Process for dewatering sludges |
| US6982040 | 2670 | 12 | 2003 | 4 | 0.1109 | 0.88889 | 7 | Method and apparatus for purifying water |
| US4026794 | 7 | 1 | 1976 | 16.858 | 0.46738 | 0.81463 | 11 | Process for resolving oil-in-water emulsions by the use of a cationic polymer and the water soluble salt of an amphoteric metal |
| US5458781 | 1608 | 6 | 1992 | 15.1449 | 0.41989 | 0.67581 | 9 | Bromide separation and concentration using semipermeable membranes |
| US6923901 | 2649 | 11 | 2002 | 5 | 0.13862 | 1 | 2 | Non-chemical water treatment method and apparatus employing ionized air purification technologies for marine application |
| US4784774 | 869 | 5 | 1987 | 29.4067 | 0.81529 | 1 | 13 | Compositions containing phosphonoalkane carboxylic acid for scale inhibition |
| US6761827 | 2542 | 11 | 2001 | 4 | 0.1109 | 0.8 | 6 | Method and apparatus for purifying water |
| US4952327 | 1059 | 6 | 1988 | 15.4514 | 0.42838 | 0.68948 | 8 | Scale control with terpolymers containing styrene sulfonic acid |
| US4387027 | 413 | 2 | 1981 | 24.619 | 0.68255 | 1 | 5 | Control of iron induced fouling in water systems |
| US7329343 | 2842 | 11 | 2005 | 5 | 0.13862 | 1 | 5 | Water treatment bypass loops having ozone and chlorine generators |
| US4454176 | 500 | 3 | 1982 | 23.3972 | 0.64868 | 0.72168 | 2 | Supported reverse osmosis membranes |
| US5603840 | 1755 | 9 | 1995 | 16.375 | 0.45399 | 0.91608 | 3 | Method of achieving microbiological control in open recirculating cooling water |
| US4234421 | 245 | 2 | 1979 | 15.8684 | 0.43994 | 0.64456 | 5 | Land restoration following oil-well drilling |
| US4952326 | 1058 | 6 | 1988 | 11.3167 | 0.31375 | 0.50498 | 6 | Dispersion of particulates in an aqueous medium |
| US7416666 | 2886 | 12 | 2003 | 4 | 0.1109 | 0.88889 | 7 | Mobile desalination plants and systems, and methods for producing desalinated water |
| US7320761 | 2839 | 14 | 2006 | 4 | 0.1109 | 1 | 4 | Method for purifying water |
| US4566972 | 620 | 3 | 1985 | 12.0925 | 0.33526 | 0.37299 | 4 | Treatment of aqueous systems |
| US5171451 | 1306 | 8 | 1991 | 16.2522 | 0.45059 | 0.60457 | 3 | Simultaneous use of water soluble polymers with ozone in cooling water systems |